\documentclass[12pt,preprint]{aastex}
\usepackage{graphicx}
\usepackage{emulateapj5}
\usepackage{apjfonts}
\usepackage{epsfig}
\usepackage{natbib}
\def\gtrsim{\mathrel{\hbox{\rlap{\hbox{\lower4pt\hbox{$\sim$}}}\hbox{\raise2pt\hbox{$>$}}}}}

\newcommand{\hst}{\emph{HST}}

\newcommand{\mbh}{\ensuremath{M_\mathrm{BH}}}

\newcommand{\msigma}{\ensuremath{M_{\mathrm{BH}}-\sigmastar}}

\newcommand{\sigmastar}{\ensuremath{\sigma_{\ast}}}

\def\lax{{$\mathrel{\hbox{\rlap{\hbox{\lower4pt\hbox{$\sim$}}}\hbox{$<$}}}$}}
\def\gax{{$\mathrel{\hbox{\rlap{\hbox{\lower4pt\hbox{$\sim$}}}\hbox{$>$}}}$}}

\begin{document}

\title{Using Megamaser Disks to Probe Black Hole Accretion}

\author{Jenny E. Greene\altaffilmark{1,8}, 
Anil Seth\altaffilmark{2}, Mark den Brok\altaffilmark{2}, 
James A. Braatz\altaffilmark{3}, Christian Henkel\altaffilmark{4,9},
Ai-Lei Sun\altaffilmark{1}, Chien Y. Peng\altaffilmark{5}, 
Cheng-Yu Kuo\altaffilmark{6}, C. M. Violette Impellizzeri\altaffilmark{3,7}, 
K. Y. Lo \altaffilmark{3}}

\altaffiltext{1}{Department of Astrophysics, Princeton University, Princeton, NJ 08540}
\altaffiltext{2}{University of Utah, Salt Lake City, UT 84112}
\altaffiltext{3}{National Radio Astronomy Observatory, 520 Edgemont Road, 
Charlottesville, VA 22903, USA}
\altaffiltext{4}{Max-Planck-Institut f{\"u}r Radioastronomie, Auf dem H{\"u}gel 69, 
53121 Bonn, Germany}
\altaffiltext{5}{GMTO Corporation.  251 S. Lake Ave., Suite 300.  Pasadena, C 91101}
\altaffiltext{6}{ASIAA}
\altaffiltext{7}{Joint Alma Office, Alsonso de Cordova 3107, Vitacura,  Santiago, Chile}
\altaffiltext{8}{Alfred P. Sloan Fellow}
\altaffiltext{9}{Astron. Dept., Kind Abdulaziz University, P.O. 80203, Jeddah, Saudi Arabia}

\begin{abstract}
  We examine the alignment between H$_2$O megamaser disks on sub-pc
  scales with circumnuclear disks and bars on $<500$ pc scales
  observed with \hst/WFC3.  The \hst\ imaging reveals young stars,
  indicating the presence of gas.  The megamaser disks are not
    well aligned with the circumnuclear bars or disks as traced by stars
    in the \hst\ images.  We speculate on the implications of the
  observed misalignments for fueling supermassive black holes in
  gas-rich spiral galaxies.  In contrast, we find a strong preference
  for the rotation axes of the megamaser disks to align with radio
  continuum jets observed on $\gtrsim 50$ pc scales, in those galaxies
  for which radio continuum detections are available.  Sub-arcsecond
  observations of molecular gas with ALMA will enable a more complete
  understanding of the interplay between circumnuclear structures.
\end{abstract}

\section{Introduction}
\label{sec:Introduction}

Active galactic nuclei have always posed a basic and fundamental
problem -- how to cram gas that is happily rotating on kpc scales onto
an accretion disk on AU scales \citep[e.g.,][]{balickheckman1982}.  We
do not know the mechanism that dissipates angular momentum and allows
gas to accrete.  There are no shortage of ideas, including major or
minor mergers \citep[e.g.,][]{sandersetal1988,hopkinsetal2006}, bars
or bars within bars
\citep[e.g.,][]{shlosmanetal1990,maciejewskietal2002,
  huntetal2008,kimetal2012}, or nuclear spirals
\citep[][]{englmaiershlosman2000,maciejewski2004,
  martinietal2003,annthakur2005}.  In a couple of nearby cases,
inflows are directly observed along circumnuclear spirals in ionized
gas on hundreds of pc scales
\citep[e.g.,][]{storchibergmannetal2007,daviesetal2009}.

\begin{figure*}
\vbox{ 
\vskip 0mm
\hskip +20mm
\includegraphics[width=0.8\textwidth]{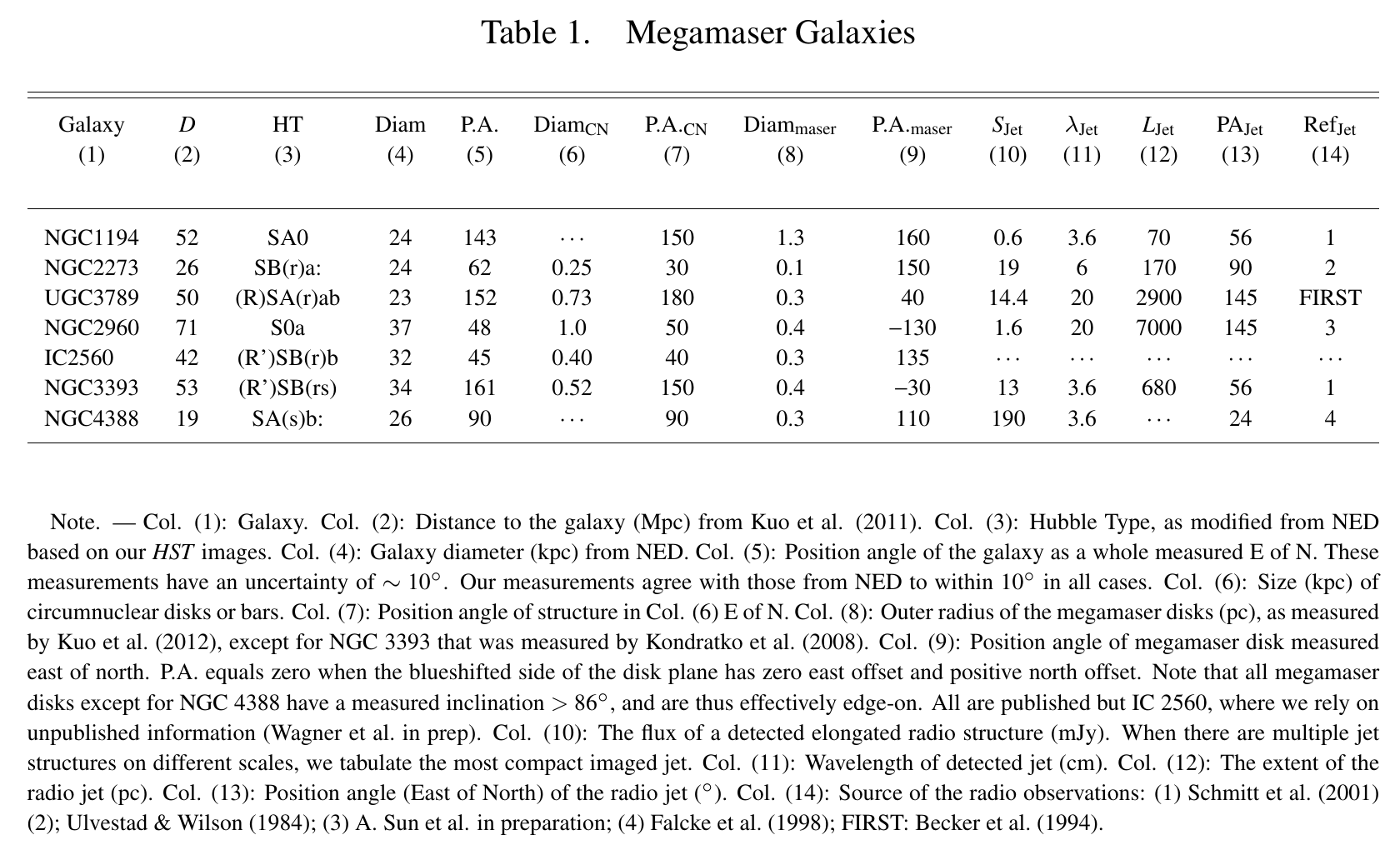}
}
\end{figure*}

One intriguing clue to the origin of accreting gas comes from the
well-known misalignment between the rotation axes of accretion disks
and the galaxy-scale disk, typically measured using radio jets as a
tracer of the angular momentum on sub-pc scales
\citep[e.g.,][]{ulvestadwilson1984}.  In this paper we use 22 GHz
water megamaser emission as a tracer of sub-pc--scale accretion disks
\citep[e.g.,][]{miyoshietal1995,lo2005}. Early examples of megamaser
disks revealed the same misalignment, this time between the outer
galaxy disk and the $\sim 0.5$ pc molecular disk producing the masers
\citep[e.g.,][]{braatzetal1997,greenhilletal2009}.  The interpretation
is ambiguous. Perhaps the galaxy swallowed some small gas-rich galaxy
that is providing fuel with a random orientation, or perhaps the accretion
process involves torques in which angular momentum is not conserved as
a function of scale.

The sample of megamaser disk galaxies mapped with Very Long Baseline 
Interferometry has grown considerably in the
past few years \citep[e.g.,][]{reidetal2009,kuoetal2011}, and the
trend of misalignment continues.  Here, we contribute \emph{Hubble
  Space Telescope} observations of nine megamaser host galaxies.
Combining \hst/WFC3 imaging from F336W to F160W, we are able to
identify disk-like structures on $\lesssim 500$ pc scales in the
majority of the sample galaxies.  Even on these circumnuclear scales, we
find no evidence for alignment with the megamaser disks.

\section{Circumnuclear Structures}
\label{sec:Observations}

We observed nine megamaser disk galaxies with \hst\ in Cycle 18 under
GO-12185.  We have examined the \msigma\ relation of these galaxies
already \citep{greeneetal2010}, and proposed to study the correlations
between galaxy bulge properties and \mbh\ with \hst. The galaxies are
all $\sim L^*$ spirals with $M_B \approx 21$ mag and Hubble types
ranging from S0 to Sbc \citep{greeneetal2010}.  In two orbits, we
obtained F336W, F438W, F814W, F110W, and F160W (roughly $UBIJH$)
images of each galaxy with integration times of $1320, 430, 2140, 150,
420$~sec respectively.  Two galaxies (NGC 6264 and NGC 6323) are more
distant than 100 Mpc, which prohibits the analysis presented here.
Thus we limit our attention to the seven nearest galaxies in our
sample (Table 1), including five galaxies from \citet{kuoetal2011},
NGC 3393 \citep{kondratkoetal2008}, and IC 2560
\citep{ishiharaetal2001}.  Published VLBI maps reveal the orientation
of the masing disk very precisely in all cases but IC 2560, where we
rely on unpublished information (Wagner et al.\ in prep).  We follow
the convention of Kuo et al.\ and refer to the position angle (PA) of
the megamaser disk as the angle East of North to the blue-shifted side
of the disk.

\subsection{Identifying circumnuclear structures}

We use a combination of ellipse-fitting and color maps to identify
organized circumnuclear structures such as nuclear bars
\citep[e.g.,][]{maciejewskietal2002,erwinsparke2003}, nuclear rings
\citep[e.g.,][]{buta1986}, or nuclear spirals
\citep[e.g.,][]{martinietal2003}.  The ellipse fits are performed on
the F160W images using the algorithm of \citet{jedrzejewski1987} as
implemented by the IRAF code {\tt ellipse}.  The code fits ellipses to
the galaxy isophotes, allowing the center, position angle (PA), and
ellipticity ($e = 1 - b/a$, where $a, b$ are the major and minor axis 
respectively) to vary as a function of radius.  Circumnuclear structures
are identified based on peaks and valleys in ellipticity with
corresponding changes in PA pointing to the presence of nuclear bars
or spirals \citep[e.g.,][]{erwinsparke2003}.  Color information
provides secondary clues as to the nature of the structures that we
identify.  For instance, nuclear rings are identified as red or blue
rings \citep[e.g.,][]{butacrocker1993}.  Note that while we do not
have direct observations of gas yet, blue colors that identify recent
star formation identify gas-rich regions by inference.  Our primary
interest here is not in identifying all galactic structures.  Instead,
we are most interested in the projected PA of the circumnuclear
($\lesssim 500$ pc) structures, for comparison with the PA of the
megamaser disks (Figure \ref{fig:disk}).

We must be careful in interpreting blue colors in the circumnuclear
region, because the active galactic nuclei (AGNs) are exciting
narrow-line region emission that also falls into our broad-band
filters. We use a pixel-by-pixel fitting method to isolate pixels that
are dominated in the blue by narrow emission lines.  For each pixel in
the three optical bands (F336W, F438W, F814W), we explore a grid of
single stellar population (SSP) models \citep[][]{bressanetal2012},
with metallicities in the range [Fe/H]$= -1$ to 0 (although our
results are not sensitive to the exact range of metallicity), and age
running from 1~Myr to 13~Gyr. Each grid point produces two colors,
effectively $U-B$ and $B-I$.  Dust is present in almost all of our
galaxies, but we are only interested in identifying emission-line
gas. Thus we use, for each pixel and grid point, a range of extinction
values (A$_V = 0-50$) to modify the SSP colors according to a standard
extinction law \citep{girardietal2008}. We then determine the maximum
likelihood value for each pixel by comparing the likelihoods of all
grid points and extinction values. We conservatively mask all pixels
with a low maximum likelihood (corresponding to $\chi^2 > 4$).  Note
that we have not masked the F160W images in making our ellipticity
profiles.  There are emission lines in these bands, but their
equivalent width is low.  As an extreme example, the famous S-shaped
emission region in NGC 3393 is detected weakly in the F160W image, but
has no impact on the measured PA, which is nearly orthogonal to the
orientation of the narrow-line region.

\subsection{Circumnuclear structures and megamaser disks}

\begin{figure*}
\vbox{ 
\vskip -19mm
\hskip -5mm
\includegraphics[width=1.1\textwidth]{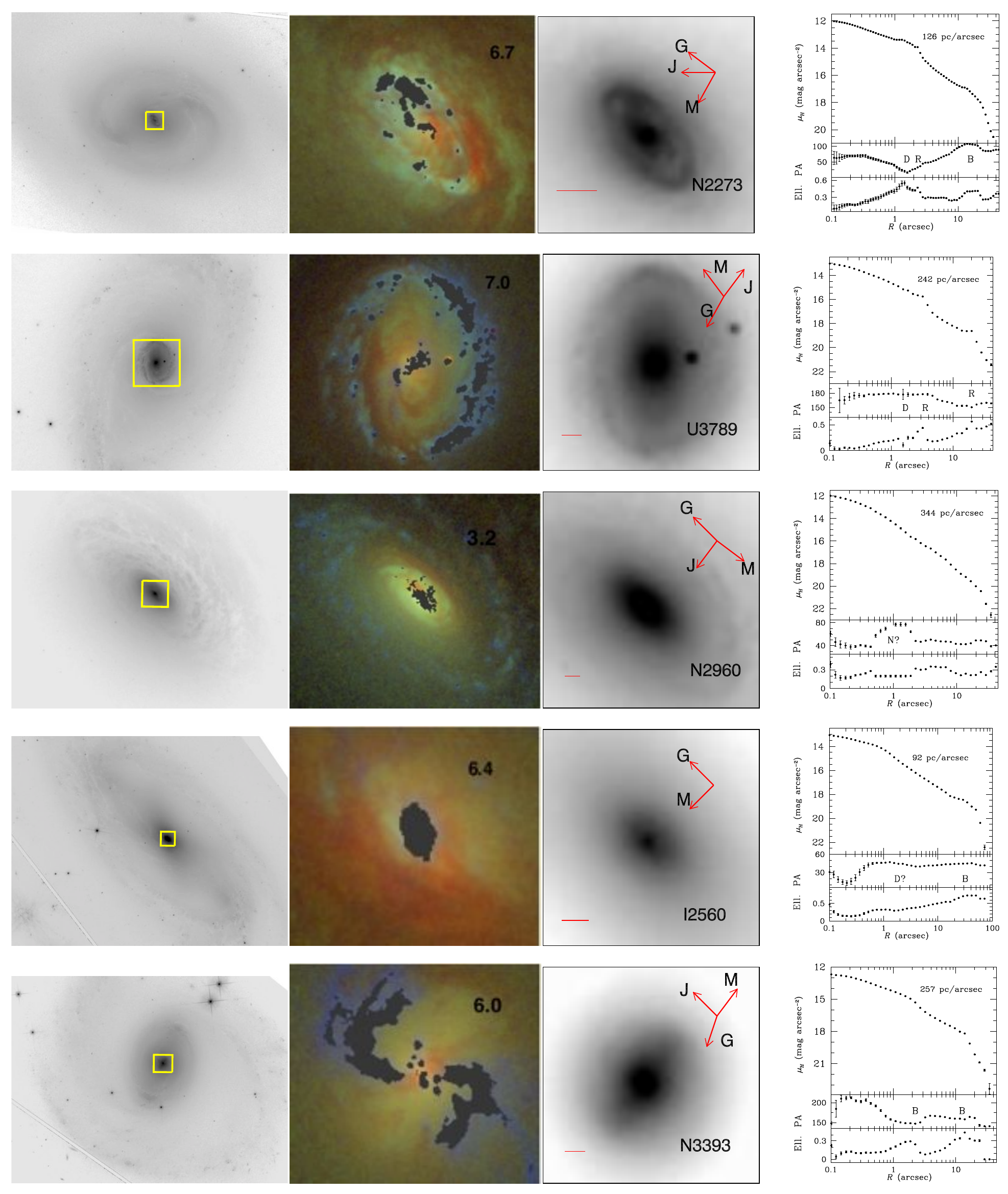}
}
\vskip -1mm
\figcaption[]{ Images of the five megamaser disk galaxies from our \hst\ program
  that appear to contain circumnuclear disks or bars on $< 500$ pc
  scales.  North is up and East to the left. For each galaxy we show,
  from left to right, the F814W image on large scales, a three-color
  image made with F336W, F438W, and F814W, and the F160W image.  The
  central and right-most images are shown on the same scale, as
  indicated by the yellow box in the left-most image. The angular scale (in arcsec) 
  is indicated in the upper right of the color image, while the red
  scale bar in the right-most image indicates 200 pc. In the central
  three-color image, red light indicates both reddened regions and
  older stellar populations, while blue light indicates recent star
  formation.  The grey
  regions are pixels that have been masked because the colors are
  dominated by narrow line emission rather than starlight.  In the
  right-most F160W image the red arrows show the orientation of the
  kpc-scale galaxy (G), the megamaser disk (M) and a jet if known (J;
  see also Table 1).  Recall that the megamaser disks are effectively
  edge-on in all cases, unlike the large-scale galaxies, and the arrow
  points towards the blue-shifted component of the disk.  Finally, we
  show the radial profile from {\it ellipse}, along with the radial
  distribution in ellipticity and PA. All identified structures are
  marked as B=bar, D=disk, N=narrow-line emission, and R=ring.
\label{fig:disk}}
\end{figure*}

In three galaxies (NGC 2273, NGC 2960 or Mrk 1419, and UGC 3789) we
see evidence for a circumnuclear disk.  In the case of NGC 2273 and
UGC 3789, there is an inner ring which manifests as a spike in the
ellipticity profile, and then dust-lanes interior to the ring most
readily identified with spiral arms.  In the case of NGC 2273,
kinematic data confirm this interpretation
\citep{barbosaetal2006,falconbarrosoetal2006}.  We find a blue
structure in the inner $\sim 3$\arcsec\ of NGC 2960 with constant PA
and ellipticity as the outer disk.  Within the inner $\sim 1$\arcsec,
there is a PA change likely associated with the narrow-line region
gas, as shown (Figure \ref{fig:disk}).

NGC 3393 also shows a bar and maybe a ring on $\sim 20$\arcsec\
scales, and then has what appears to be a nuclear bar on $\sim
3$\arcsec\ scales.  At the position of the bar, we see an ellipticity
maximum at a constant position angle, the classic signature of a bar
\citep[e.g.,][]{wozniaketal1995,menendez-delmestreetal2007}.  IC 2560
is a difficult case, because the central blue light is dominated by
narrow-line emission.  On kpc-scales ($>30$\arcsec), the galaxy shows
the classic signature of an X-shaped bar \citep[e.g.,][]{lietal2011}.
However, in the nucleus, dust obscuration and narrow-line emission
conspire to make identification of any organized structure quite
difficult.  Since the PA profile is flat for radii $r > 0.5$\arcsec,
we assign IC 2560 an inner PA of 45\degr, but caution that the
identification in this case is uncertain. Finally, we have no
information for NGC 4388 or NGC 1194, which are edge-on, such that
dust from the galaxies on large scales obscures the nucleus.  In these
two cases we adopt the PA of the large-scale disk (Figure
\ref{fig:edgeon}).

\vskip +1mm
\hskip -10mm
\includegraphics[width=0.52\textwidth]{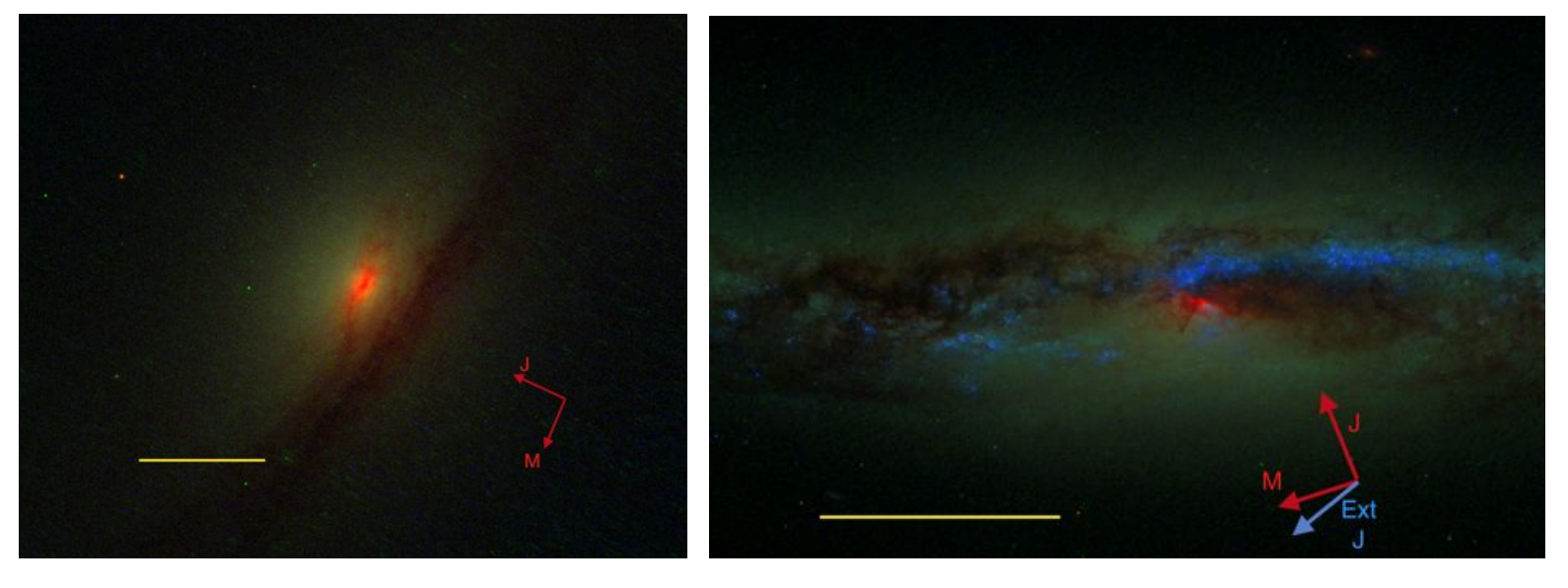}
\vskip -0mm
\figcaption[]{Three color images for the edge-on galaxies NGC 1194 (left) and 
NGC 4388 (right), constructed as above but without masking.  
The yellow scale bar indicates 1 kpc 
while the red arrows indicate the jet (J) and the direction towards the 
blue-shifted side of the megamaser disk (M).
The blue arrow indicates the extended jet (Ext J) directions.
\label{fig:edgeon}}
\vskip 5mm

\begin{figure*}
\vbox{ 
\vskip -19mm
\hskip +25mm
\includegraphics[width=0.75\textwidth]{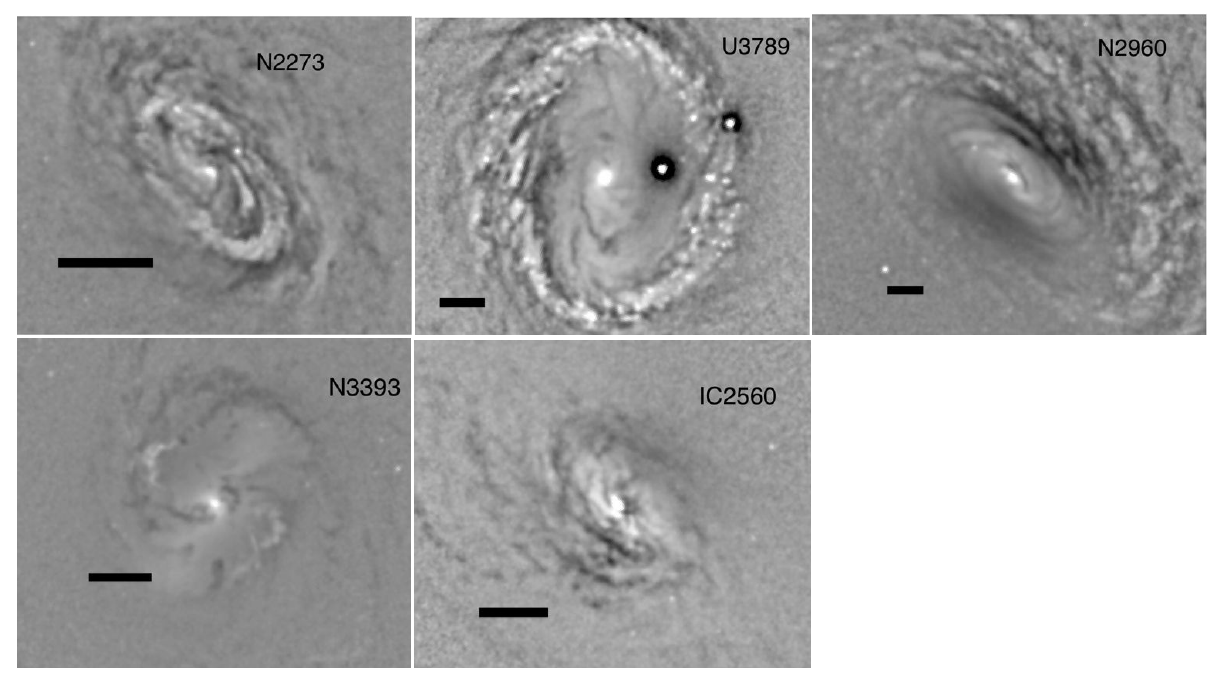}
}
\vskip -1mm
\figcaption[]{Structure maps \citep{poggemartini2002}
made in with F814W to emphasize the circumnuclear dust structure. 
Dark regions are dust lanes while bright regions correspond to emission line 
knots due to either star formation or narrow-line region gas.  We have zoomed in 
on comparable regions as seen in the color maps in Figure \ref{fig:disk}, with 
the black scale bars again indicating 200 pc.
In the case of NGC 2273, NGC 2960, and IC 2560, the dust structure is clear.  UGC 3789 
appears to show two arms reaching towards the center inside the inner ring.  The structure 
in NGC 3393 is dominated by the narrow-line gas.
\label{fig:struct}}
\end{figure*}

As an alternate way to look at the data, we also generate structure
maps from the F814W images \citep{poggemartini2002} shown in Figure
\ref{fig:struct}.  Structure maps use deconvolution to remove
large-scale power and highlight fine structures from dust (dark in our
maps) and emission-line regions excited either by the AGN or star
formation (bright in the structure maps). These emphasize the dust
structures in NGC 2273, NGC 2960, and IC 2560.  There is dust
structure apparent in UGC 3789, although it is subtle.  However, the
structure in NGC 3393 is dominated by the narrow-line region emission
(see also the mask in Figure \ref{fig:disk}).  In summary, for the
five megamaser disk galaxies studied here, whose large-scale galaxies
are not edge-on (Figure \ref{fig:disk}), we find clear evidence for
circumnuclear disks or bars in four cases.  We previously suggested
that the majority of the megamaser disks are found in pseudobulge
galaxies \citep{greeneetal2010}.  Now, with the high spatial
resolution of \hst, we have confirmed that there is evidence for
ongoing secular evolution in the circumnuclear regions of these
galaxies, as expected for pseudobulges
\citep[e.g.,][]{kormendykennicutt2004}.

Having identified these circumnuclear structures, we now must assign a
projected position angle to each.  We use the PA from the ellipse
fitting at the scale of the feature, often corresponding to a local
extreme value in ellipticity. As described in \citet{erwinsparke2003},
projection effects (e.g., the superposition of a bright bulge and a
nuclear bar) can lead to offsets between the isophotes and the
measured PA.  However, based on visual inspection of our galaxies, we
believe that our values are good to $\sim 10$\degr.  Table 1 contains
the orientation and linear size of each circumnuclear structure.  The nuclear
disk in NGC 2273 (radius of 2\arcsec\ or 260 pc) is well-studied
\citep{erwinsparke2003}, and our PA agrees with the literature value
within our uncertainty of 10\degr.  For the remainder of the paper, we
will focus on comparisons between the PA of these circumnuclear
structures on 100-500 pc scales with the megamaser disks on sub-pc
scales.  If we need to refer to larger scales in the galaxy, we will 
refer to the kpc-scale galaxy.

In principle, we are also interested in the distribution of
inclinations for these circumnuclear structures.  Taking the
ellipticity measurements at the same radius as the PA measurements and
assuming thin disks, we find that all of the targets have high
inclinations ($i > 55$\degr). However, it is not at all clear that we
are identifying thin disks in all cases.

We show the distribution of $\Delta$PA$\equiv$PA$_{\rm disk} -
$PA$_{\rm maser}$ in Figure \ref{fig:delpa}{\it a}.  The first main
result of this paper, highlighted in Figure \ref{fig:delpa}{\it a}, is
that the megamaser disks do not tend to align with the circumnuclear
structures ($\lesssim$ 500 pc) traced by \hst.  We have also included
NGC 1068, Circinus, and NGC 3079, where the circumnuclear disk PA is
measured from molecular disks. In the case of Circinus, the megamaser
disk is warped, and we take the PA of the inner disk.  The molecular
disk in NGC 1068 is warped, but in inclination rather than in PA. 

Of course, we are interested in the orientation of the stars and the
gas.  With the \hst\ images, we infer the presence of gas based on
either blue colors corresponding to recent star formation or dust
lanes.  We have neither gas or stellar kinematic observations at these
sub-arcsecond scales yet.  However, in the megamaser galaxies NGC
1068, NGC 3079, Circinus, and NGC 2273, detailed gas observations do
exist.  The detailed interpretation is different in each case.  In NGC
1068, \citet{schinnereretal2000} probe $\sim 20$ pc scales with CO,
and show that the molecular gas disk has a warp that starts at $\sim
70$~pc and extends all the way to the maser disk on 1~pc scales. NGC
3079 contains a kpc-scale molecular disk that is aligned with the
major axis of the galaxy, and then a 600-pc scale molecular oval that
is aligned in PA both with the kpc-scale disk and the pc-scale masing
disk \citep{kodaetal2002,kondratkoetal2005}.  In Circinus, the maser
disk is nearly perpendicular to the bipolar radio jet and a CO outflow
\citep{greenhilletal2003}.  Finally, the kinematic major and minor
axes of the disk in NGC 2273 are not aligned, and the disk is likely
warped \citep[e.g.,][]{barbosaetal2006}. The CO disk appears to align
with the warped disk on 200 pc scales \citep{petitpaswilson2002},
while the megamaser disk is misaligned in PA by $60\degr$.  Existing
observations of gas and kinematics thus suggest that the circumnuclear
structures observed here are manifestations of torques that facilitate
accretion onto smaller scales.

In conclusion, for objects where $\lesssim 500$pc-scale gas or stellar
structure observations are available, $40\%$ show a misalignment
$\gtrsim 20\degr$ with the masing disks.  We see no tendency for
alignment between the stellar structures traced by \hst\ and the
megamaser disks, although the gas structures do tend to align.  Note
that because we focus here on projected position angles, we are
observing a lower limit on the misalignments.

\section{Nuclear Jet to Disk Angle}
\label{sec:Jet}

We have seen that stellar structures on $\lesssim 500$ pc are
frequently misaligned with the sub-pc scale accretion disk.  We still
expect that the nuclear jet should emerge along the rotation axis of
the sub-pc scale disk \citep[e.g.,][]{pringleetal1999}.  Since the
megamaser disks must be close to edge-on in order for amplification to
occur, the jets should be perpendicular to the megamaser disk
positions on the sky. Indeed, \citet{henkeletal2005} found a high rate
of megamaser detections in galaxies where the jet was likely in the
plane of the sky, suggesting that jets are usually aligned with the
rotation axis of the maser disk.  We are able to observe the relative
orientation of the megamaser disk rotation axis and the jet directly.

Most objects in our sample have been observed in the radio continuum
at relatively high angular resolution (Table 1), including NGC 1194,
NGC 3393 \citep{schmittetal2001}, NGC 2273 \citep{ulvestadwilson1984},
IC 2560 \citep{morgantietal1999}, and NGC 4388 \citep{falckeetal1998}.
In the case of UGC 3789, we simply quote the FIRST fit on kpc scales
since no higher resolution data are yet available.  Finally, for NGC
2960 we measured a marginally resolved structure at 20 cm from our
EVLA observations with a size of $20 \pm 3$\arcsec\ (A. Sun et al. in
preparation).  Only IC 2560 is unresolved (on the 1\arcsec\ or 0.2 kpc
scales measured by Morganti et al.).  We caution that these
observations probe jets on tens of pc scales and in some cases, VLBI
observations do reveal PA changes on pc scales
\citep[e.g.,][]{ulvestadetal1998,middelbergetal2004}.  In fact,
  the pc-scale jet in NGC 3079 is misaligned \citep{trotteretal1998},
  while the diffuse radio emission on pc scales is aligned with the
  rotation axis of the disk \citep{duricseaquist1988}. Given the
alignment that we report on tens of pc scales, it would be quite
interesting to see different behavior on pc scales for more of our
sources, in light of the possibility that the jet axis may be
  determined on small scales by the black hole spin but on large (tens
  of pc scales) by electromagnetric forces from the disk
  \citep{mckinneyetal2013}.

\begin{figure*}
\vbox{ 
\vskip -3mm
\hskip +20mm
\includegraphics[width=0.75\textwidth,angle=0]{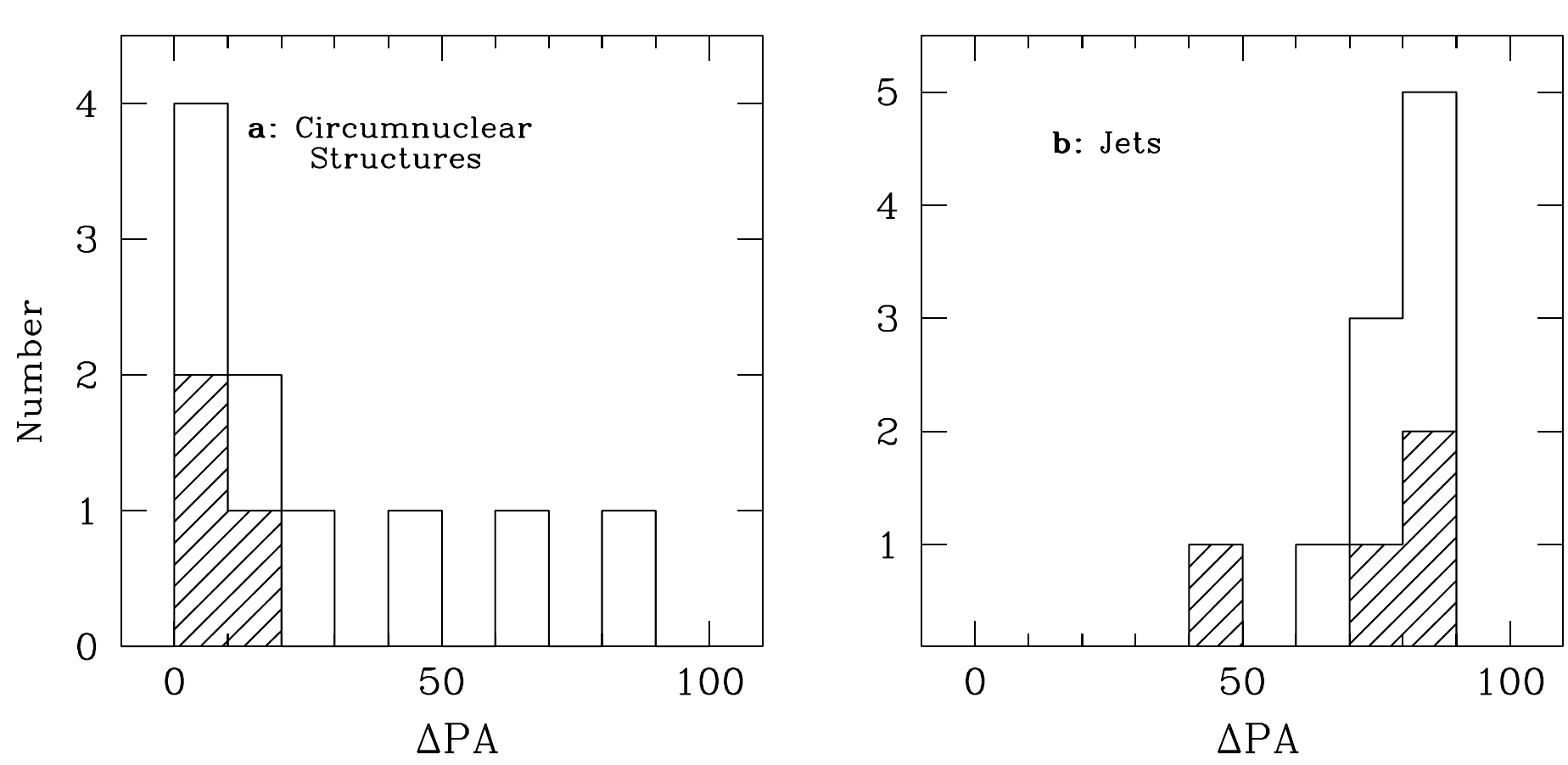}
}
\vskip -0mm
\figcaption[]{
Histogram of the difference in projected position angle between the megamaser disk 
and ({\it left}): the circumnuclear structures on $\lesssim$500 pc scales identified in the 
\hst\ images and ({\it right}): the most compact jet structure we could 
find in the literature, as summarized in Table 1
(note -- this is the PA of the megamaser disk, 90\degr\ from the rotation axis).
{\it Left}: In the case of NGC 1194 and NGC 4388 we simply use
the large-scale PA of the galaxy. 
The filled histograms indicate the literature megamasers, where we use  
the difference in projected position angle between 
known CO disks in NGC 1068 \citep{schinnereretal2000}, 
Circinus \citep{curranetal1998}, and NGC 3079 \citep{kondratkoetal2005}. 
Thus, these measurements are not strictly comparable (ours probes recent 
star formation while the other directly probes molecular gas).
{\it Right}: In this case the filled histograms include 
NGC 4258 as well.  
\label{fig:delpa}}
\end{figure*}

In Figure \ref{fig:delpa}{\it b}, we show the distribution of
$\Delta$PA for the projected jet and megamaser disk angles.  Note
that the megamaser disks are edge-on, so the PA is determined by
measuring the angle on the sky of the observed linear alignment of
maser spots and is 90\degr\ from the rotation axis of the disk.  As
Figure \ref{fig:delpa} shows, the vast majority of the jets are
aligned with the rotation axis of the megamaser disk, as expected.
We more than double the number of systems with a direct disk-jet
comparison \citep{herrnsteinetal1999,
  greenhilletal2003,kondratkoetal2005}. The largest outliers are are
NGC 3079 ($\Delta$PA$=44$\degr) and NGC 2273
($\Delta$PA$=60$\degr). In summary, in all but one of the systems,
the jets on tens of pc to kpc scales are found to align within
$<15\degr$ of the megamaser rotation axis.

In addition to compact jets, we would eventually like to investigate
possible kpc-scales jets to seek evidence of a change in direction of
the accretion disk on longer timescales.  For instance,
radiation-pressure--driven warping may drive sub-pc scale disk
precession on $\sim 10^6$ year timescales
\citep[e.g.,][]{pringle1997}. Furthermore, it is interesting to ask
whether ionization cones align with the jet/megamaser disk rotation
axis.  We do not really have complete information on the orientation
of the narrow-line regions for the full sample.  However, from the
masked (grey) pixels in Figure \ref{fig:disk}, we can see that the
spectacular S-shaped narrow-line region in NGC 3393 does align with
the radio jet and the megamaser rotation axis
\citep[e.g.,][]{cookeetal2000,kondratkoetal2008}, and this is seen in
Circinus as well \citep{greenhilletal2003}.  In other cases, we do not
have adequate information yet to address the relative orientation of any 
ionization cones.

\section{Warps and Kinematic Misalignments Facilitate Accretion}

\vskip 5mm

We have seen that nuclear (and in some cases extended) radio jets
align with the rotation axis of the megamaser disks. In contrast, the
sub-pc scale accretion disk does not align with the kpc-scale galactic
disk (see \S\ref{sec:Introduction}).  Using \hst/WFC3 imaging, we have
found no strong tendency for alignment between the megamaser disks and
the circumnuclear structures traced by stars on $\lesssim 500$ pc
scales.  Here we review what these misalignments may tell us about the
accretion process.

{\it Disk warping on pc scales causes the observed misalignments.}
One route to disk misalignment is warping of the pc-scale accretion
disk. According to \citet[e.g.,][]{pringle1997}, the sub-pc scale disk may
change position angle by a significant fraction due to radiation
pressure.  For the megamaser disks, the typical timescale for disk PA
changes due to radiation-pressure warping would be $\sim 5 \times
10^6$~yr \citep[see also][]{maloneyetal1996,gammieetal2000} with only
a linear dependence on BH mass. AGN lifetimes are very uncertain, but
\citet{martinietal2003} suggest that accretion episodes in
low-luminosity sources may only last millions of years, perhaps just
long enough for some PA change over the AGN lifetime.  Note that
radiation pressure warping cannot cause the warps observed on
circumnuclear ($\sim 200$ pc) scales in NGC 2273 and NGC 1068.

Another possible source of sub-pc--scale torques comes from stars in a
cusp around the BH; \citet{bregmanalexander2012} find that resonant
relaxation with a stellar cusp will drive warping of order 10\degr\ on
sub-pc scales, comparable to the warp in NGC 4258.  The expected
timescale for evolution is $\sim 10^7$ yr for NGC 4258, perhaps in
some tension with the million-year lifetimes from Martini et al.\
Alternatively, \citet{kartjeetal1999} suggest that masing clumps are
accelerated out of the disk plane by magnetic pressure to a height
where the balance of shielding and pumping is optimal.  Since the
optimal height will depend on the AGN luminosity, they predict more
warping at higher luminosities. Both radiation pressure and stellar
resonance warping act on sub-pc scales, i.e.\ on the scales of the
megamaser disks.  Warping does not appear to be a compelling
explanation in general for the observed misalignments for three
reasons.  First, in contrast to NGC 4258, the observed warps in the
new megamaser disks are $<10$\degr\ 
\citep[][]{impellizzerietal2012,reidetal2013,kuoetal2013} and thus smaller than
predicted by either warping model.  Second, the observed warps are too
small to explain the broad range of PA differences that we observe.
Third, the timescales for these warps to operate may be long compared
with typical lifetimes.

{\it Changes in angular momentum as a function of scale result
  naturally in accretion events.}  \citet{hopkinsquataert2010} study
the progression of gas from galaxy-wide to pc scales.  Because the gas
is dissipational, the kinematics of the gas and stars decouple. As a
result, the stars are able to torque the gas, leading to warps and
ultimately shifts in the angular momentum vector of the gas as a
function of scale.  Observations of the gas kinematics on $\sim 100$
pc scales in nearby active galaxies reveal that fueling is indeed
facilitated by decoupled dynamical components such as inner bars,
ovals or spirals \citep[e.g.,][]{huntetal2008,garciaburilloetal2009},
but that these fueling episodes are stochastic and short-lived
\citep[e.g.,][]{dumasetal2007,haanetal2009}, in keeping with the observed kinematic
misalignment on all spatial scales that we can probe.

In the Hopkins \& Quataert models a massive bulge component will
suppress these torques, thus requiring different feeding mechanisms
(e.g., merging) in bulge-dominated systems \citep[see also their
analytic model in ][]{hopkinsquataert2011}.  It is intriguing to note
that the megamaser aligns with the large-scale disk in the two
bulge-dominated galaxies NGC 1194 and NGC 2960.  Our statistics 
are obviously quite limited.  We can use our strong
confirmation that nuclear jets align with the rotation axis of
megamaser disks (and by extension sub-pc scale disks in general). We
look for morphology-dependent alignment in the \citet{kinneyetal2000}
Seyfert sample, with morphological types from the RC3.  There is no
preference for jet alignment with the rotation axis of the kpc-scale
galaxy disk among S0 or S0/a galaxies.  However, it would be
interesting to investigate possible jet alignment with the rotation
axes of circumnuclear $\sim 500$ pc-scale structures in the Kinney et
al.\ sample (e.g., using \hst).

{\it Disk misalignments boost accretion rate.}  \citet{nixonetal2012}
show that if there are strong misalignments between inner and outer
disks, then the gas angular momentum is dissipated where the disks
meet, facilitating accretion.  As a result, {\it if} there is a major
misalignment as a function of scale, then the accretion rate onto the
BH can be boosted by an order of magnitude.  We may preferentially
observe AGN activity at times of misalignment.  With a much larger 
sample, we could look for an anti-correlation between AGN luminosity and 
PA alignment as predicted by this model.

{\it The gas has an external origin, and thus knows nothing about the
  angular momentum of the disk.}  It would be very challenging to
completely rule out that the gas is supplied at random angles via
accretion of small satellites. In fact, in the case of NGC 3393 there
is tentative evidence for an accretion event in the putative detection
of two AGN separated by $\sim 150$ pc in projection
\citep{fabbianoetal2011}.  Nevertheless, none of the targets shows
evidence for morphological disturbance.  Given the high incidence of
circumnuclear disks and bars in the sample, an internal origin for the
accreted gas, with accretion facilitated by secular processes, seems
more natural for the bulk of the sample
\citep[e.g.,][]{kormendykennicutt2004}.

\section{Summary}

Using \hst/WFC3 observations of megamaser disk galaxies, we have shown
that the majority of the megamaser disk galaxies contain clear
evidence for circumnuclear structures (disks or bars) often with
associated star formation.  We find that the accretion disk on sub-pc
scales (as traced by the megamaser disks) shows no strong tendency to 
align with circumnuclear (often star-forming) structures
such as bars, spirals, or rings on $<500$ pc scales.  In contrast, we
find that the rotation axis of the megamaser disk is usually aligned with the
compact radio jet axis.  We review possible explanations for the
misalignment between circumnuclear and sub-pc scale disks, and favor a
scenario in which accretion naturally requires changes in angular
momentum as a function of scale.  
Given the good correlation between the jet and megamaser
rotation axis, in the future we can boost the sample statistics using
\hst\ observations of Seyfert galaxies with radio jets.  Furthermore,
future direct observations of circumnuclear gas in the megamaser disk
galaxies (e.g., with ALMA) will help clarify the processes that feed
the central monster.

\acknowledgements We would like to thank Paul Martini, Jeremy Goodman,
Eliot Quataert, Jim Ulvestad, and John Kormendy for very useful
conversations. We thank the anonymous referee for a timely and
thorough report that improved this manuscript.

\end{document}